\documentclass[a4paper,pra,superscriptaddress,11pt,showkeys]{revtex4}
\usepackage[a4paper,centering,hmargin=1.75cm,vmargin={2cm,3.6cm}]{geometry}
\usepackage{pdfpages}
\usepackage{pgffor}

\usepackage{graphicx}

\usepackage{amsmath}
\usepackage{amsfonts}
\usepackage{amssymb}

\begin{document}

\title{Low-confinement silicon nitride waveguides manufactured via direct glass bonding}

\author{Mikhail~V.~Tsvetkov}
\email{mikesportfun56@gmail.com}
\affiliation{Institute of Nanotechnology of Microelectronics of the Russian Academy of Sciences, Moscow 119334, Russia}

\author{Dmitry~V.~Obydennov}
\email{obydennovdv@my.msu.ru}
\affiliation{Institute of Nanotechnology of Microelectronics of the Russian Academy of Sciences, Moscow 119334, Russia}

\author{Alexandr~S.~Rykov}
\affiliation{Institute of Nanotechnology of Microelectronics of the Russian Academy of Sciences, Moscow 119334, Russia}

\author{Alexandr~R.~Shevchenko}
\affiliation{Institute of Nanotechnology of Microelectronics of the Russian Academy of Sciences, Moscow 119334, Russia}

\author{Maxim~V.~Shibalov}
\affiliation{Institute of Nanotechnology of Microelectronics of the Russian Academy of Sciences, Moscow 119334, Russia}

\author{Ivan~A.~Filippov}
\affiliation{Institute of Nanotechnology of Microelectronics of the Russian Academy of Sciences, Moscow 119334, Russia}

\author{Stepan~D.~Perov}
\affiliation{Institute of Nanotechnology of Microelectronics of the Russian Academy of Sciences, Moscow 119334, Russia}

\author{Nikita Yu. Dmitriev}
\affiliation{Institute of Nanotechnology of Microelectronics of the Russian Academy of Sciences, Moscow 119334, Russia}

\author{Michael A. Tarkhov}
\affiliation{Institute of Nanotechnology of Microelectronics of the Russian Academy of Sciences, Moscow 119334, Russia}

\begin{abstract}
Reducing the fabrication cost of photonic integrated circuits while maintaining low optical losses and technological simplicity is essential for their wider implementation. In conventional manufacturing methods, the dielectric cladding thickness around waveguides is usually limited to $\sim20$~$\mu$m, which complicates suppression of radiative losses and parasitic scattering in low-confinement geometries. In this paper, we propose and experimentally demonstrate an alternative technology for forming low-confinement waveguides in Borofloat~33 glass by thermal fusion bonding of two glass wafers. The waveguide pattern is formed by etching trenches with depths on the order of tens of nanometers into the glass, filling them with silicon nitride, removing the excess layer, and bonding the planarized glass surfaces, thereby forming a thick, symmetric dielectric cladding. As a proof of concept, we fabricated straight waveguides with a core height of 50~nm and widths from 1.3 to 3.5~$\mu$m. With butt coupling to standard SMF-28 single-mode fiber at 1550~nm,  we obtained chip transmissions up to 60\%, corresponding to input/output coupling losses of $\sim1$~dB per facet and consistent with numerical estimates. Fabry--Perot analysis of high-resolution spectra measured with AR-coated lensed fibers gave effective propagation losses down to $0.62\pm0.36$ dB/cm, depending on waveguide width and polarization. The proposed approach provides a simple and scalable route to low-confinement glass-encapsulated photonic circuits with passive butt coupling, promising for long delay lines, external-cavity laser feedback circuits, and ring-resonator sensors.
\end{abstract}

\keywords{bonding, planar technology, integrated photonics, silicon nitride, low confinement, fiber-to-chip coupling}

\maketitle

\section*{Introduction}
Silicon-nitride-based integral photonics has become one of the key areas for application systems in the telecom range due to the combination of wide spectral transparency~\cite{munoz_silicon_2017,buzaverov_silicon_2024}, CMOS-compatibility~\cite{moss_new_2013, romero-garcia_silicon_2013} and  potentially low propagation losses~\cite{Bauters2011SiN,el_dirani_ultralow-loss_2019,liu_high-yield_2021,pfeiffer_ultra-smooth_2018}. However, optimal waveguide geometry  (in particular, the degree of mode localization)  is determined by the trade-offs between losses, nonlinearity,  thermal stability, bending radius, and fiber interface requirements. In this work, we focus on low-confinement SiN$_x$-waveguides, which are often implemented through an ultra-thin core and/or a large effective mode area.

The primary engineering advantage of low-confinement geometries is the reduced losses due to sidewall roughness and fabrication fluctuations: a smaller fraction of the field is concentrated at the core-cladding interface, which weakens scattering on microscopic defects and facilitates  the achievement of ultralow losses under the same lithography and etching constraints. Early demonstrations of low-loss  silicon nitride waveguides relied precisely on field delocalization and geometry optimization to minimize scattering, achieving loss levels on the order of several dB/m or below in the telecom window~\cite{Tien2010,Bauters2011SiN}. The second practical advantage is the larger effective mode area and, consequently, lower optical intensity at the same power, which increases the maximum power and reduces the impact of Kerr nonlinearity which is important for delay lines, high dynamic range filters, and external laser resonators~\cite{Tien2010}. The third advantage is improved fiber coupling: a mode size close to that of the fiber enables reduced input/output losses and simplifies passive packaging, which is critical for application-specific circuits. Finally, low-confinement waveguides are often preferable for applications where key metrics are phase noise/intensity noise (via high-Q resonators and long delay lines) and scaling stability, including coherent telecom systems and microwave photonics~\cite{Roeloffzen2013,Xiang2021}.

Weak confinement increases the minimal bend radii (often to the centimeter scale), thereby increasing chip area and element footprint. Additionally, it reduces the flexibility of dispersion engineering and integration density compared to high-confinement platforms, where smaller bends are achievable at the cost of stricter requirements for roughness and width/height variations.

From a practical standpoint,  low-confinement platforms impose several requirements. First, material absorption must be minimized, including absorption associated with N--H and Si--H bonds and control of composition/defects, which typically requires high-temperature annealing.~\cite{Tien2010,Ji2021} Second, the ultrathin silicon nitride thickness must be controlled with high wafer-scale uniformity, because even small size variations can significantly alter the effective refractive index and phase delays in long structures. Third, stress and cracking must be controlled, since silicon nitride exhibits high intrinsic stress; known approaches include specialized patterns for crack stopping and stress partitioning.~\cite{Luke2013Stress,Ji2021} Also, a high-quality lower/upper SiO$_2$ cladding (low roughness, low impurities) is required for low losses, along with minimization of optical interaction with potentially defective interfaces.

The most common  techniques for fabricating such structures involve LPCVD deposition of silicon nitride followed by etching and SiO$_2$ cladding formation; the choice of waveguide geometry (thicknesses and widths) enables trade-offs between confinement, bend losses, and scattering~\cite{Bauters2011SiN,Tien2010}. To improve yield and control cracking in ``thick''  silicon nitride (often associated with high-confinement but technologically relevant for hybrid platforms as well), the Damascene process is widely used, where the waveguide is formed by filling pre-etched trenches followed by planarization, ensuring controlled stress and high Q-factors in microrings~\cite{Pfeiffer2016Damascene}. For long delay lines and high-Q resonators, polishing/planarization techniques (e.g., chemical-mechanical polishing (CMP)), optimization of plasma etching, and sidewall micro-roughness reduction are also critical~\cite{Ji2021}.

Fabrication methods that place the waveguide pattern between two high-quality SiO$_2$ layers are of special interest for low-confinement waveguides: they produce symmetric cladding, robust mechanical encapsulation, and minimized optical losses. In integrated photonics, this is typically achieved using oxide-to-oxide direct bonding (hydrophilic/fusion bonding) or plasma-activated direct bonding to reduce annealing temperature and time. Adhesive bonding (e.g. BCB) and metal-mediated methods are also used when material compatibility outweighs the need for ultralow optical losses~\cite{Xu2020BondingReview}. In photonic platforms, oxide-oxide bonding is a basic operation for multilayer integration (layer transfer and vertical transition formation), while maintaining low losses in waveguide regions~\cite{Bauters2013SiOnSiN}. In the context of ``two glass wafers'', notable examples include wafer-bonded structures with thermally grown SiO$_2$ as high-quality cladding, demonstrating record-low propagation losses and detailed analysis of surface cleanliness/planarity requirements and annealing conditions~\cite{Bauters2011WaferBonding}.

In practical implementation for  silicon nitride chips, the typical process sequence usually includes formation of the lower SiO$_2$ layer (thermal oxide or PECVD/HDPCVD followed by planarization), deposition and patterning of SiN$_x$, formation of the upper SiO$_2$ layer with surface treatment (CMP if necessary), surface activation (plasma/UV-ozone or chemical cleaning), and alignment and direct bonding, followed by low-/high-temperature annealing to enhance bond strength and optical homogeneity~\cite{Xu2020BondingReview,Bauters2013SiOnSiN}. This approach yields a mechanically and thermally stable ``sandwich'' structure, convenient for further edge processing, integration with other layers, and wafer-scale scaling.

For fabricating test samples of waveguide structures, it is worth mentioning femtosecond laser writing (FSLW) technology~\cite{PhysRevApplied.22.064079}. This technology enables the creation of waveguide structures in various types of glass: amorphous silicon oxide~\cite{Shah:05}, borosilicate glass~\cite{chen_femtosecond-laser-written_2018}, phosphate glass~\cite{Dong:13}, and various glass alloys~\cite{10.1117/1.1905363}. 
However, scaling this technology to industrial production is challenging.

More broadly, glass-based photonics has recently developed beyond femtosecond-laser-written structures. Recent examples include low-loss polymeric waveguides on glass substrates for co-packaged optics~\cite{Chang2025PolymerGlass}, silicon nitride photonics on glass for scalable optical redistribution layers~\cite{Jin2025SiNGlass}, ion-exchanged glass waveguide circuits for high-density optical interconnects~\cite{Brusberg2025IonExchangedGlass}, and low-confinement TiO$_2$ waveguides fabricated by electron-beam evaporation~\cite{Hsu2025TiO2}. These approaches illustrate that glass-based platforms can also be compatible with scalable manufacturing. In this context, the proposed method is not intended to replace state-of-the-art LPCVD or Damascene SiN technologies, which provide lower losses, mature stress-control strategies, and higher integration density. Its main distinction is the use of direct glass bonding to form a thick, symmetric, mechanically robust cladding around an ultrathin SiN$_x$ core, while retaining a simple route to low-confinement geometries and efficient butt coupling.

In this work, we present an alternative technology for obtaining low-confinement waveguides in Borofloat 33 glass, utilizing thermal bonding to fuse two glass surfaces. This approach allows reducing the minimum waveguide thickness to tens of nanometers, enables low-loss input from standard optical fiber ($\sim 1$ dB per facet), and offers scalability for manufacturing. The proposed process is potentially compatible with wafer-level scaling because the waveguides are defined by parallel lithographic, deposition, planarization, and bonding steps rather than by serial writing; however, scaling to larger wafers will require dedicated control of surface roughness, flatness, wafer bow, CMP uniformity, particle contamination, bonding yield, and chip-level facet preparation.

\section*{Chip manufacturing}

\begin{figure}[htb!]
    \centering
    \includegraphics[width=\textwidth]{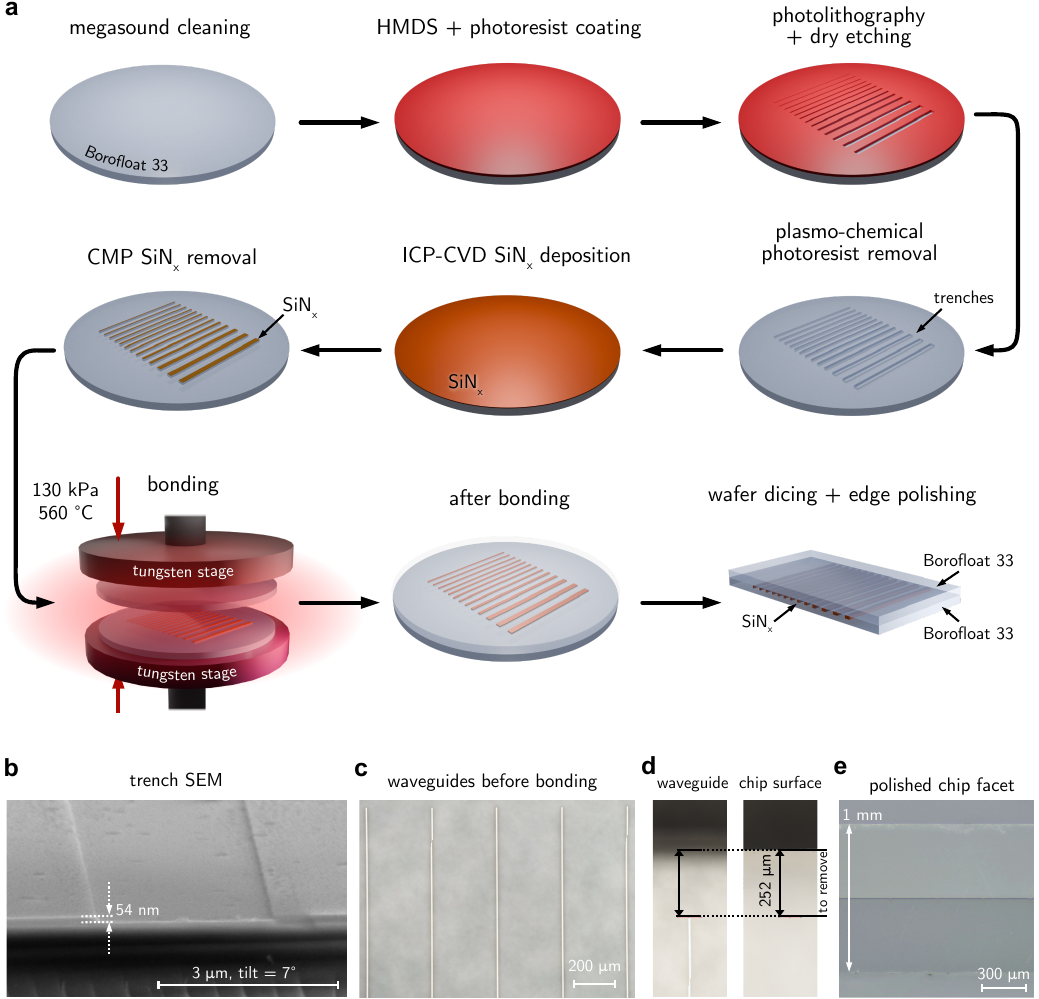}
    \caption{[Picture modified] Fabrication of low-confinement waveguide. a. Schematic of optical waveguide fabrication stages, including wafer preparation, application of photoresist and etching of Borofloat 33 wafer, silicon nitride deposition and processing, bonding, dicing, and chip polishing. HMDS—hexamethyldisilazane, ICP-CVD --- inductively coupled plasmo-chemical vapour deposition. b. SEM image of a wafer cross-section with an etched trench. The measured trench depth of 54 nm is  shown. c. Optical image of fabricated  silicon nitride waveguides before bonding. d. Image of a diced chip edge showing the distance subsequently removed during facet polishing. Image is focused on the waveguide plane and the chip top surface. e. Optical image of the polished chip facet.}
    \label{pipeline}
\end{figure}

The fabrication of low-confinement SiN$_x$ waveguides with thermal bonding technology included a few stages, as illustrated in Figure~\ref{pipeline}a: Borofloat 33 wafer megasound cleaning; hexamethyldisilazane (HMDS) and photoresist coating;  waveguides structures photolitography; trenches dry etching; plasma cleaning glass wafer of the remaining photoresist; silicon nitride ICP-CVD deposition  (as described elsewhere~\cite{mumlyakov_void-free_2024}); remaining SiN$_x$ layer  removing by chemical-mechanical polishing; thermal glass-glass bonding under force; wafer dicing and mechanical edge polishing.

In this work, we used Borofloat~33 glass (500 $\mu$m-thick 100 mm plates). Its advantages over quartz glass include lower cost and the ability for thermal fusion bonding due to the low softening temperature of 820~$^{\circ}$C~\cite{Borofloat33}. Silicon nitride was selected as the waveguide material, as it is already employed in low-confinement technologies. To demonstrate the feasibility of the proposed method, we fabricated and characterized an array of waveguides with widths in the range of 1.3--3.5 $\mu$m and a height of approximately 50 nm. The modal composition expected for this thickness and width range is analyzed in Section~I of the Supporting Information.

 Successful glass--glass bonding requires both surfaces to be flat. To satisfy this requirement, we formed trenches in the glass substrate according to the waveguide topology and then filled them with silicon nitride. After removal of the excess silicon nitride, the wafer surface became planar and suitable for bonding.

For etching, a 740 nm layer of photoresist (Microposit AZ1505) was applied to the surface of one of the wafers with HMDS as a primer, followed by UV lithography of the waveguide topology. After photoresist development, $\sim55$~nm trenches were dry etched in the glass wafer; one of etched trench is presented on SEM image (Figure~\ref{pipeline}b). The remaining photoresist was removed then by plasma cleaning and megasonic washing.

Next, a 150 nm thick SiN$_x$ layer was deposited onto the glass wafer using inductively coupled plasmo-chemical vapour deposition (ICP-CVD) with monosilane and nitrogen gases. The remaining SiN$_x$ was removed by chemical-mechanical polishing, resulting in flat glass surface and SiN$_x$-filled  trenches.

Images of the fabricated waveguides after removal of excessive silicon nitride are shown in Figure~\ref{pipeline}c. After that, the wafer surfaces were megasonically cleaned again, and immediately after that, the two surfaces were mechanically pressed together. Thermal bonding was performed using an AML AWB 4 system. After achieving vacuum in the chamber, heating began at a rate of 2~$^{\circ}$C/min to 560~$^{\circ}$C, and the wafers were pressed together with a force of 1000~N, corresponding to a pressure of 130 kPa. At these conditions, the wafers were held for 1 hour and then  the pressed wafers were allowed to cool naturally inside the bonding chamber to room temperature ($\sim$23~$^\circ$C) over approximately 8~h.

After that, chip dicing was performed. To minimize input coupling losses, each chip's facets were polished to remove all material up to the waveguide start; the distance from the chip edge to the waveguides was measured using optical microscopy (Figure~\ref{pipeline}d). Polishing was performed using diamond film. An optical photograph of the chip facet after polishing is shown in Figure~\ref{pipeline}e.

\section*{Optical Measurements}
\begin{figure}[htb!]
    \centering
    \includegraphics[width=\textwidth]{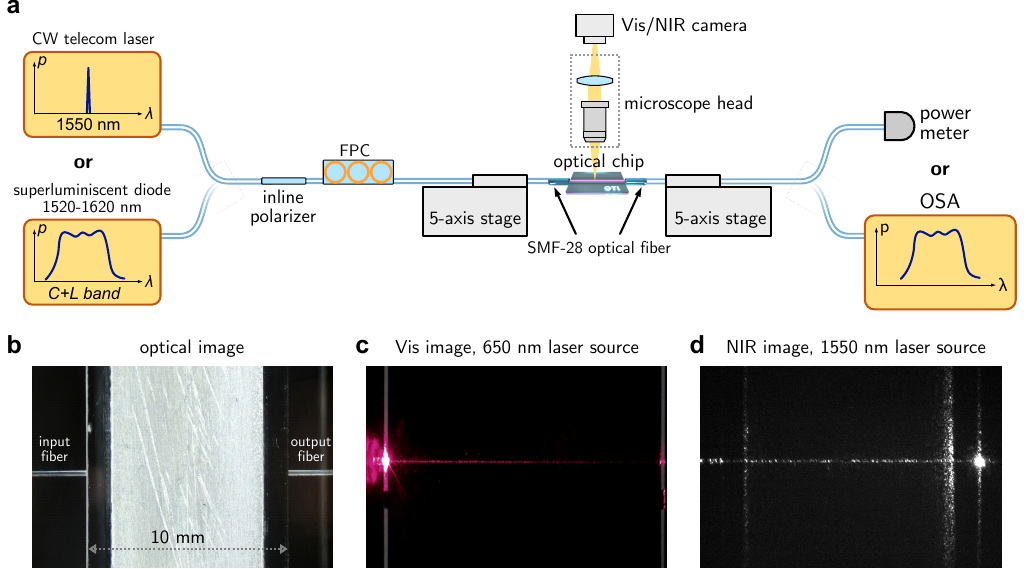}
    \caption{Waveguide transmission measurement setup. a. Setup schematic. FPC --- fiber polarization controller. b. Optical image of the chip with waveguides and coupled input/output fibers. Waveguide length (10 mm) is indicated in Figure. c. Image of visible test source (650 nm) propagation through the waveguide. d. Image of 1550 nm laser radiation propagation through the waveguide.}
    \label{setup}
\end{figure}

We measured waveguide transmission using butt-coupling with standard single-mode cleaved SMF-28 fiber. The setup schematic is shown in Figure~\ref{setup}a. We used a telecom-band diode laser with a wavelength of 1550 nm. Polarization control was achieved using a fiber polarizer and polarization controller (FPC). Input and output fibers were mounted on 5-axis stages, enabling precise butt-coupling of fiber facets to the chip's waveguide facets, as shown in Figure~\ref{setup}b. Fiber position alignment was performed by maximizing the output optical power. The output  power $P_\text{out}$ was  measured using a Thorlabs PM 400 power meter . 

Waveguide transmission was calculated as $T = P_\text{out}/P_\text{in}$. For spectral measurements, we used a broadband superluminescent diode source covering a range of 1520--1620 nm, and an optical spectrum analyzer (Ceyear 6362D). The transmission spectrum was evaluated as $T(\lambda) = I_\text{out}(\lambda)/I_\text{in}(\lambda)$, where $I_\text{in}$ and $I_\text{out}(\lambda)$  are the input and output spectra, measured at the input and output fibers, respectively.

For visual monitoring of light coupling, we used a visible or NIR camera attached to a microscope. A top-view image of the chip with input and output fibers is shown in Figure~\ref{setup}b. Laser radiation propagation (650 nm) is shown in Figure.~\ref{setup}c. NIR laser radiation propagation  (1550 nm) is shown in Figure~\ref{setup}d. One can see that over the waveguide length scale of 1 cm, no visible attenuation of scattered NIR radiation is observed, which may indicate relatively low propagation losses. 

\begin{figure}[htb!]
    \centering
    \includegraphics[width=\textwidth]{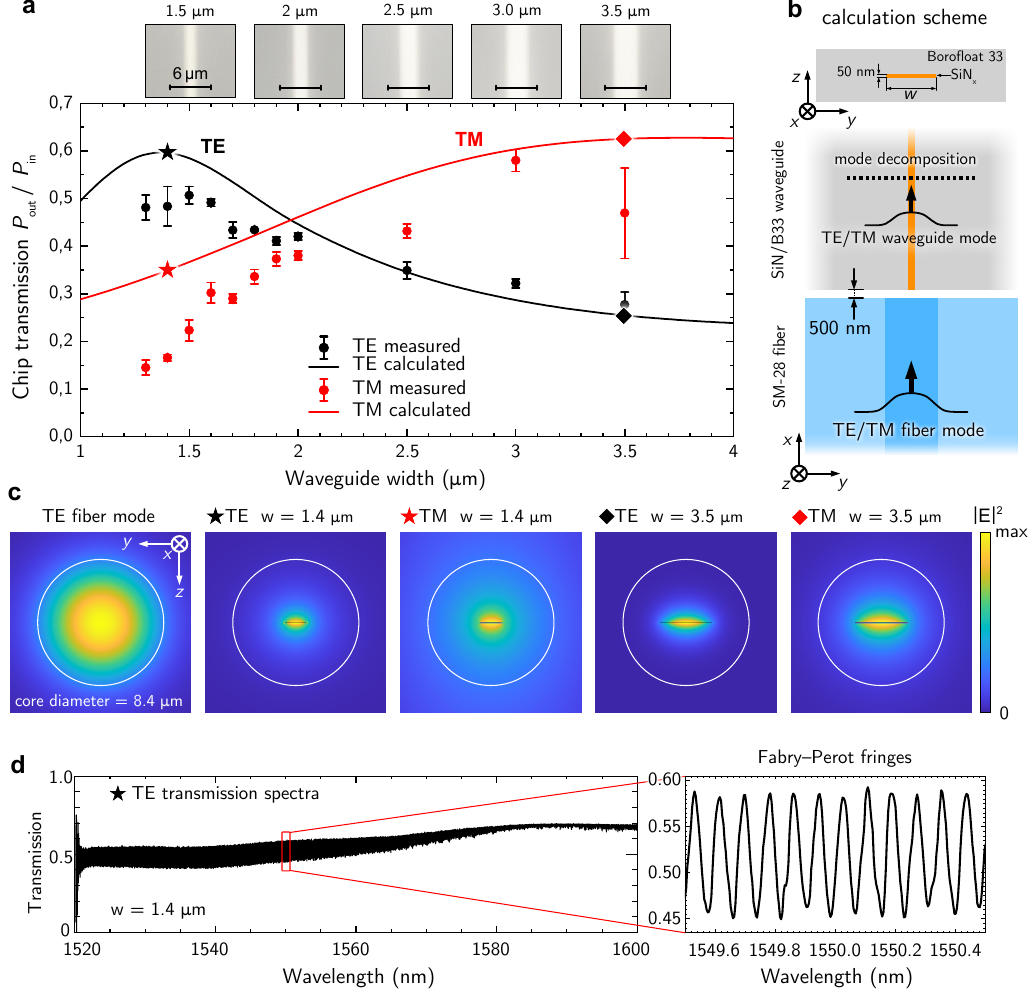}
    \caption{Waveguide transmission dependence on waveguide width. a. Waveguide transmission vs. width. Black dots correspond to measured transmission values for polarization along the y-axis, red dots—along the z-axis. Black and red curves represent calculated transmission for TE and TM modes, respectively. Top insets show optical images of waveguides with measured widths indicated. b. Coupling calculation schematic with axis orientations. c. Squared magnitude of electric field distribution in the SMF-28 fiber TE mode, TE and TM modes of 1.4 $\mu$m-wide waveguide (marked by black and red stars on the graph), and TE and TM modes of 3.5 $\mu$m-wide waveguide (marked by black and red diamonds). Square side length of presented distributions is 12 $\mu$m; white circle denotes the fiber core area with 8.4 $\mu$m diameter. d. The measured transmission spectrum for a 1.4 $\mu$m-wide waveguide and TE polarization; left panel --- full spectrum, right panel --- the area near 1550 nm showing Fabry--Perot fringes.}
    \label{width}
\end{figure}

Results of waveguide transmission measurements as a function of their width are presented in Figure~\ref{width}.
 Within the studied waveguide width range (1.3--3.5 $\mu$m), there are two propagating modes, TE and TM, differing in radiation polarization. For separate characterization of the transmission of the two modes, the input polarization was selected either in the plane of planarization (y-axis in Figure~\ref{width}), corresponding to the TE mode, or in the perpendicular plane (z-axis), corresponding to the TM mode. Each measurement was repeated for three waveguides of identical width. Figure~\ref{width}a shows the results averaged over three measurements; error bars correspond to the standard error of the mean. For a waveguide width of 1.4~$\mu$m and TE polarization, we also measured the transmission spectrum (Figure~\ref{width}d). The spectrum exhibits pronounced Fabry--Perot fringes caused by reflections from the chip facets.

A maximum transmission of $\sim50$\% was achieved for waveguide widths of 1.3--1.5 $\mu$m with TE polarization, and approximately 50--60\% for widths of 3--3.5 $\mu$m with TM polarization. In general, waveguide transmission consists of three components:
$
T = T_\text{in} T_\text{prop} T_\text{out},
$
where $T_\text{in}$ and $T_\text{out}$ --- coupling coefficients from fiber to waveguide and from waveguide to fiber, respectively, which are determined by butt-coupling losses, $T_\text{prop}$ describes propagation losses in the waveguide. With symmetric input/output coupling, by the reciprocity principle, it can be assumed that $T_\text{in} \approx T_\text{out}$, and therefore $T \approx T_\text{in}^2 T_\text{prop}$. This allows us to estimate the input/output losses as $T_\text{in/out} \geq \sqrt{T}$. For TE polarization, $T_\text{TE} \geq 70\%$, and for TM polarization, $T_\text{TE} \geq 77\%$.

The primary factor affecting the optical power coupling coefficient is the field overlap between the source and the waveguide mode. In low-confinement waveguides the mode size increases as waveguide dimensions decrease~\cite{Bauters2011SiN}, so there should be an optimal size at which the fiber-waveguide mode overlap is maximized. To verify this, we performed three-dimensional simulations of light coupling from a single-mode fiber into the waveguide using the finite-difference time-domain (FDTD) method to account for Fresnel reflection effects at interfaces. The simulation schematic is shown in Figure~\ref{width}b; typical SMF-28 single-mode fiber parameters were used, with the fundamental TE mode at 1550~nm selected as the source (polarized along the y-axis for TE-mode transmission calculations or z-axis for TM mode). The refractive index of Borofloat~33 was set to $n_\mathrm{B33}=1.4557$ at 1550~nm~\cite{polyanskiy_refractiveindexinfo_2024}, while the refractive index of the deposited SiN$_x$ film was measured by ellipsometry as $n_\mathrm{SiN}=2.01$ at 1550~nm. The gap between the waveguide edge and fiber was set to 500~nm. Radiation transmitted through the 50-nm-thick waveguide with variable width was analyzed using a mode decomposition method, which allowed us to compute the fraction of intensity corresponding to the desired mode, yielding the transmission coefficient $T_\text{in}$; Figure~\ref{width}a shows the computed $T_\text{in}^2$ values as curves. As one can see, the calculated dependencies qualitatively match the measurements, though measured values are systematically lower. This can be attributed to propagation losses in fabricated waveguides, particularly for widths below 2 $\mu$m; here, weak mode confinement may enhance coupling to free space and adjacent waveguides, thereby increasing propagation losses.

For illustration, at the points of maximum calculated transmission (for $w=1.4~\mu$m and $w=3.5~\mu$m),Figure~\ref{width}c shows the TE and TM mode profiles compared to the fiber mode. For $w=1.4~\mu$m, the TE mode exhibits the best overlap with the fiber mode, while the TM mode shows the opposite  because of weak confinement. For $w=3.5~\mu$m, the effective mode sizes decrease: the TE mode becomes smaller than optimal, whereas the TM mode achieves optimal overlap with the fiber mode.

The difference between measured and loss-free calculated chip transmission is on the order of tens of percent. Given the 10~mm length of the fabricated waveguides, this difference indicates propagation losses on the order of 1~dB/cm. To refine this estimate, we additionally evaluated propagation losses from Fabry--Perot oscillations in the transmission spectra, following the standard approach based on the Airy response of a low-finesse waveguide cavity~\cite{Regener1985,Walker1985}. Since the extracted loss is sensitive to the effective facet reflectivity, special attention was paid to reducing parasitic feedback from the measurement fibers.

\begin{figure}[htb!]
    \centering
    \includegraphics[width=\textwidth]{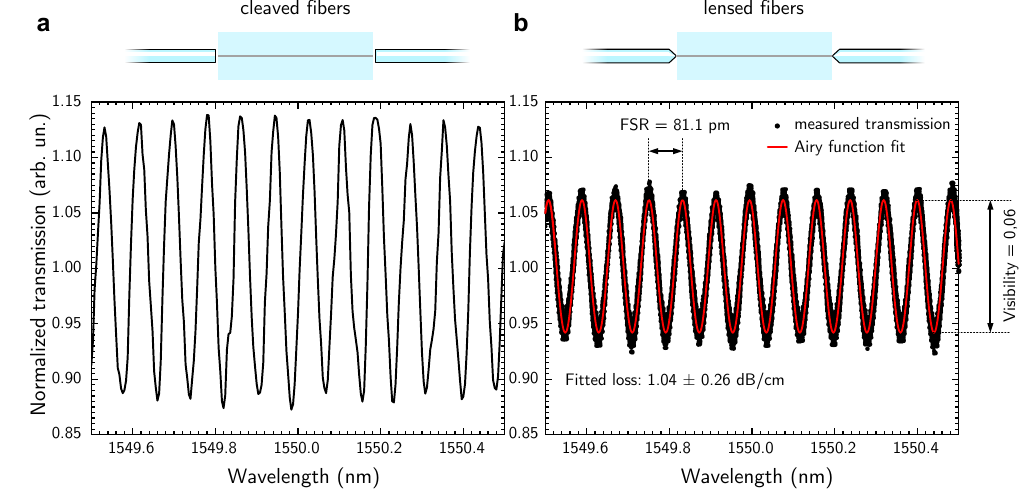}
    \caption{Waveguide transmission spectra measured under different fiber-coupling conditions. a. Normalized transmission measured using a pair of 90$^\circ$-cleaved SMF-28 fibers. The increased fringe visibility indicates additional reflection from the fiber facets. b. Normalized transmission measured using AR-coated lensed fibers. Black dots correspond to the measured spectrum, and the red curve shows the Airy-function fit used for Fabry--Perot loss estimation. }
    \label{spectra}
\end{figure}

For this purpose, we compared spectra measured with cleaved SMF-28 fibers and with AR-coated lensed fibers (Raysung Photonics, 90$^\circ$ cone angle, AR coating with $R<0.5\%$ at 1550~nm, and 40--50$^\circ$ full divergence angle, corresponding to $\mathrm{NA}\approx0.34$--0.42 in air). The cleaved fibers produced higher Fabry--Perot visibility (Figure~\ref{spectra}a), indicating additional reflection from the fiber facets and therefore larger uncertainty in the effective cavity reflectivity. The lensed fibers reduced the fringe visibility (Figure~\ref{spectra}b), and these spectra were used for the loss extraction.

To avoid undersampling of the Fabry--Perot fringes, we recorded high-resolution scans with a tunable laser. The wavelength axis was calibrated using a pre-calibrated Mach--Zehnder interferometer, following the approach described in Ref.~\cite{Dmitriev2022JETP}. The spectra were fitted by the Airy transmission function
\begin{equation}
T_\mathrm{FP}(\lambda)=A(\lambda)
\frac{(1-R)^2\tau}
{1+R^2\tau^2-2R\tau\cos\varphi(\lambda)},
\end{equation}
where $A(\lambda)$ is a slowly varying spectral envelope, $R$ is the effective facet reflectivity, $\tau=10^{-\alpha L/10}$ is the single-pass internal transmission of the waveguide, $\alpha$ is the propagation loss in dB/cm, and $L=1$~cm is the waveguide length. Fresnel estimates and numerical calculations gave $R\approx0.034$ for the waveguide facet; however, using this value led to systematically underestimated losses. The best agreement with the measured spectra was obtained for an effective reflectivity of $R=0.038$. The difference of 0.004 may originate from residual reflection of the lensed-fiber facets and waveguide-facet roughness. Therefore, the extracted values should be considered as effective propagation-loss estimates rather than as an independent cut-back measurement.

Using $R=0.038$, the estimated losses were $1.04\pm0.26$~dB/cm for the 1.4~$\mu$m-wide TE mode, $1.90\pm1.35$~dB/cm for the 1.4~$\mu$m-wide TM mode, $0.65\pm0.59$~dB/cm for the 3.5~$\mu$m-wide TE mode, and $0.62\pm0.36$~dB/cm for the 3.5~$\mu$m-wide TM mode. These values are consistent with the order-of-magnitude estimate obtained from the difference between measured transmission and loss-free coupling simulations. More accurate separation of propagation loss from residual facet and coupling effects requires ring-resonator Q-factor measurements~\cite{Xiao2007Losses,Little1997Microring} or optical frequency-domain reflectometry/correlation-frequency-domain reflectometry (OFDR/CFDR)~\cite{Tokushima2022OFDR,Glombitza1993CFDR}.

Thus, both direct transmission measurements and Fabry--Perot analysis indicate propagation losses at the sub-dB/cm to few-dB/cm level, depending on waveguide width and polarization. For the spectra shown in Figure~\ref{spectra}, the measured Fabry--Perot free spectral range was 81.1~pm near 1550~nm, corresponding to a group index of
$n_g=\lambda^2/(2L\Delta\lambda)\approx1.48$
for a 1~cm-long Fabry--Perot cavity. This value is close to the refractive index of the glass cladding, as expected for a weakly confined mode with a large fraction of the optical field outside the ultrathin SiN$_x$ core. 

The present loss level is still higher than that of mature LPCVD and Damascene SiN platforms and is likely limited by fabrication-induced scattering, residual SiN$_x$ absorption, and, at lower loss levels, intrinsic Borofloat~33 absorption. Therefore, the immediate advantage of the proposed approach is not record-low propagation loss or high integration density. Rather, the platform is attractive where thick symmetric glass encapsulation, passive butt coupling, and low-confinement mode engineering are important. In sensing configurations, the weak localization of low-confinement modes can be beneficial, because a larger fraction of the optical field extends outside the core and can interact with the surrounding medium. Conversely, for phase-sensitive passive devices, this same environmental sensitivity must be suppressed; in such cases, a thick and stable glass cladding helps reduce uncontrolled perturbations. At the same time, the enlarged mode field relaxes fiber-chip alignment requirements compared with strongly confined submicron waveguides. These properties make the approach relevant for  long delay lines and external-cavity laser feedback circuits, where future device-level validation should include bending-loss simulations, ring-resonator Q-factor measurements, environmental-sensitivity analysis, and alignment-tolerance measurements. A preliminary bending-loss estimate is provided in Section~II of the Supporting Information. In sensing applications, the weakly confined mode may be useful primarily for refractive-index or concentration sensing in specially designed analyte-accessible regions, whereas temperature sensing is not the main target because the large modal fraction in glass reduces the effective thermo-optic response compared with more strongly confined SiN waveguides.

\section*{Conclusion}
In this work, a technology for fabricating low-confinement SiN$_x$ waveguides in Borofloat~33 glass is proposed and experimentally demonstrated. It is based on forming nano-trenches in a glass substrate, followed by SiN$_x$ filling, excess layer removal, and thermal bonding of two glass wafers. This method ensures high-quality contact surfaces and forms a symmetric dielectric cladding around the waveguide, which is essential for achieving low losses and mechanical stability. Test samples feature 1~cm long straight waveguides with core thicknesses of approximately 50~nm and widths of 1.3--3.5~$\mu$m. Butt-coupling to standard single-mode fiber SMF-28 at a wavelength of 1550~nm yielded transmissions up to 60\% (depending on width and polarization), corresponding to input/output losses of $\sim 1$~dB per facet.

Comparison of experiments with FDTD simulations of coupling confirms that the observed transmission dependence on waveguide width is primarily governed by modal overlap between the fiber mode and the waveguide's TE/TM modes. The systematic reduction in measured values relative to simulations indicates contributions from propagation losses and/or fabrication imperfections (facet quality, roughness, parasitic free-space coupling and cross-talks). Additional Fabry--Perot analysis of high-resolution spectra measured with AR-coated lensed fibers gave effective propagation losses of $1.04\pm0.26$~dB/cm for the 1.4~$\mu$m-wide TE mode and down to $0.62\pm0.36$~dB/cm for the 3.5~$\mu$m-wide TM mode. Because this method remains sensitive to the effective facet reflectivity, future work will include ring-resonator validation and/or OFDR/CFDR measurements for independent propagation-loss extraction. The proposed technological platform is promising for scalable manufacturing of  low-confinement circuits with simplified passive packaging, as well as devices based on long delay lines and high-Q resonators.

\section*{Declaration of competing interest}
The authors declare no conflict of interest.

\section*{Acknowledgements}
The study has been carried out with the support of project No. 125020501540-9 of the Ministry of Education and Science of the Russian Federation. Fabrication and technology characterization were carried out at large scale facility complex for heterogeneous integration technologies and silicon + carbon nanotechnologies.

\section*{Supporting Information}
The Supporting Information further details the calculated modal composition of the fabricated low-confinement waveguides and provides a bending-loss estimate for representative curved geometries.

\bibliographystyle{ieeetr} 
\bibliography{bib}

@article{Tien2010,
  author  = {Tien, Ming-Chun and Bauters, Jared F. and Heck, Martijn J. R. and Blumenthal, Daniel J. and Bowers, John E.},
  title   = {Ultra-low loss Si3N4 waveguides with low nonlinearity and high power handling capability},
  journal = {Optics Express},
  year    = {2010},
  volume  = {18},
  number  = {23},
  pages   = {23562--23568},
  doi     = {10.1364/OE.18.023562}
}

@article{Bauters2011SiN,
  author  = {Bauters, Jared F. and Heck, Martijn J. R. and John, Demis and Dai, Daoxin and Tien, Ming-Chun and Barton, Jonathon S. and Leinse, Arne and Heideman, Ren{\'e} G. and Blumenthal, Daniel J. and Bowers, John E.},
  title   = {Ultra-low-loss high-aspect-ratio Si3N4 waveguides},
  journal = {Optics Express},
  year    = {2011},
  volume  = {19},
  number  = {4},
  pages   = {3163--3174},
  doi     = {10.1364/OE.19.003163}
}

@article{Luke2013Stress,
  author  = {Luke, Kevin and Dutt, Avik and Poitras, Carl B. and Lipson, Michal},
  title   = {Overcoming Si3N4 film stress limitations for high quality factor ring resonators},
  journal = {Optics Express},
  year    = {2013},
  volume  = {21},
  number  = {19},
  pages   = {22829--22833},
  doi     = {10.1364/OE.21.022829}
}

@article{Roeloffzen2013,
  author  = {Roeloffzen, C. G. H. and others},
  title   = {Silicon nitride microwave photonic circuits},
  journal = {Optics Express},
  year    = {2013},
  volume  = {21},
  number  = {19},
  pages   = {22937--22961},
  doi     = {10.1364/OE.21.022937}
}

@article{Pfeiffer2016Damascene,
  author  = {Pfeiffer, Martin H. P. and Kordts, Arne and Brasch, Valerie and Zervas, Michail and Geiselmann, Michael and Jost, Julian D. and Kippenberg, Tobias J.},
  title   = {Photonic Damascene process for integrated high-Q microresonator based nonlinear photonics},
  journal = {Optica},
  year    = {2016},
  volume  = {3},
  number  = {1},
  pages   = {20--25},
  doi     = {10.1364/OPTICA.3.000020}
}

@article{Ji2021,
  author  = {Ji, Xingchen and Roberts, Samantha P. and Corato-Zanarella, Mateus and Lipson, Michal},
  title   = {Methods to achieve ultra-high quality factor silicon nitride resonators},
  journal = {APL Photonics},
  year    = {2021},
  volume  = {6},
  number  = {7},
  pages   = {071101},
  doi     = {10.1063/5.0057881}
}

@article{Xiang2021,
  author  = {Xiang, Cheng and others},
  title   = {High-performance lasers for fully integrated silicon nitride photonics},
  journal = {Nature Communications},
  year    = {2021},
  volume  = {12},
  pages   = {6650},
  doi     = {10.1038/s41467-021-26804-9}
}

@article{Bauters2013SiOnSiN,
  author  = {Bauters, Jared F. and Davenport, Michael L. and Heck, Martijn J. R. and Doylend, J. K. and Chen, Arnold and Fang, Alexander W. and Bowers, John E.},
  title   = {Silicon on ultra-low-loss waveguide photonic integration platform},
  journal = {Optics Express},
  year    = {2013},
  volume  = {21},
  number  = {1},
  pages   = {544--555},
  doi     = {10.1364/OE.21.000544}
}

@article{Bauters2011WaferBonding,
  author  = {Bauters, Jared F. and Heck, Martijn J. R. and John, Demis D. and Barton, Jonathon S. and Bruinink, Christiaan M. and Leinse, Arne and Heideman, Ren{\'e} G. and Blumenthal, Daniel J. and Bowers, John E.},
  title   = {Planar waveguides with less than 0.1 dB/m propagation loss fabricated with wafer bonding},
  journal = {Optics Express},
  year    = {2011},
  volume  = {19},
  number  = {24},
  pages   = {24090--24101},
  doi     = {10.1364/OE.19.024090}
}

@article{Xu2020BondingReview,
  author  = {Xu, Jikai and Du, Yu and Tian, Yezhe and Wang, Chao},
  title   = {Progress in wafer bonding technology towards MEMS, high-power electronics, optoelectronics, and optofluidics},
  journal = {International Journal of Optomechatronics},
  year    = {2020},
  volume  = {14},
  number  = {1},
  pages   = {94--118},
  doi     = {10.1080/15599612.2020.1857890}
}

@article{PhysRevApplied.22.064079,
  title = {Femtosecond-laser-written low-loss multiscan waveguides in fused silica},
  author = {Skryabin, N.N. and Zhuravitskii, S.A. and Dyakonov, I.V. and Straupe, S.S. and Kalinkin, A.A. and Kulik, S.P.},
  journal = {Phys. Rev. Appl.},
  volume = {22},
  issue = {6},
  pages = {064079},
  numpages = {13},
  year = {2024},
  month = {Dec},
  publisher = {American Physical Society},
  doi = {10.1103/PhysRevApplied.22.064079},
  url = {https://link.aps.org/doi/10.1103/PhysRevApplied.22.064079}
}

@article{moss_new_2013,
	title = {New {CMOS}-compatible platforms based on silicon nitride and {Hydex} for nonlinear optics},
	volume = {7},
	copyright = {http://www.springer.com/tdm},
	issn = {1749-4885, 1749-4893},
	url = {https://www.nature.com/articles/nphoton.2013.183},
	doi = {10.1038/nphoton.2013.183},
	language = {en},
	number = {8},
	urldate = {2026-01-29},
	journal = {Nature Photonics},
	author = {Moss, David J. and Morandotti, Roberto and Gaeta, Alexander L. and Lipson, Michal},
	month = aug,
	year = {2013},
	pages = {597--607},
}

@article{romero-garcia_silicon_2013,
	title = {Silicon nitride {CMOS}-compatible platform for integrated photonics applications at visible wavelengths},
	volume = {21},
	copyright = {https://doi.org/10.1364/OA\_License\_v1\#VOR-OA},
	issn = {1094-4087},
	url = {https://opg.optica.org/oe/abstract.cfm?uri=oe-21-12-14036},
	doi = {10.1364/OE.21.014036},
	language = {en},
	number = {12},
	urldate = {2026-01-29},
	journal = {Optics Express},
	author = {Romero-García, Sebastian and Merget, Florian and Zhong, Frank and Finkelstein, Hod and Witzens, Jeremy},
	month = jun,
	year = {2013},
	pages = {14036},
}

@article{munoz_silicon_2017,
	title = {Silicon {Nitride} {Photonic} {Integration} {Platforms} for {Visible}, {Near}-{Infrared} and {Mid}-{Infrared} {Applications}},
	volume = {17},
	issn = {1424-8220},
	url = {https://www.mdpi.com/1424-8220/17/9/2088},
	doi = {10.3390/s17092088},
	language = {en},
	number = {9},
	urldate = {2026-01-29},
	journal = {Sensors},
	author = {Muñoz, Pascual and Micó, Gloria and Bru, Luis and Pastor, Daniel and Pérez, Daniel and Doménech, José and Fernández, Juan and Baños, Rocío and Gargallo, Bernardo and Alemany, Rubén and Sánchez, Ana and Cirera, Josep and Mas, Roser and Domínguez, Carlos},
	month = sep,
	year = {2017},
	pages = {2088},
}

@article{buzaverov_silicon_2024,
	title = {Silicon {Nitride} {Integrated} {Photonics} from {Visible} to {Mid}‐{Infrared} {Spectra}},
	volume = {18},
	issn = {1863-8880, 1863-8899},
	url = {https://onlinelibrary.wiley.com/doi/10.1002/lpor.202400508},
	doi = {10.1002/lpor.202400508},
	language = {en},
	number = {12},
	urldate = {2026-01-29},
	journal = {Laser \& Photonics Reviews},
	author = {Buzaverov, Kirill A. and Baburin, Aleksandr S. and Sergeev, Evgeny V. and Avdeev, Sergey S. and Lotkov, Evgeniy S. and Bukatin, Sergey V. and Stepanov, Ilya A. and Kramarenko, Aleksey B. and Amiraslanov, Ali Sh. and Kushnev, Danil V. and Ryzhikov, Ilya A. and Rodionov, Ilya A.},
	month = dec,
	year = {2024},
	pages = {2400508},
}

@article{el_dirani_ultralow-loss_2019,
	title = {Ultralow-loss tightly confining {Si}$_{\textrm{3}}$ {N}$_{\textrm{4}}$ waveguides and high-{Q} microresonators},
	volume = {27},
	issn = {1094-4087},
	url = {https://opg.optica.org/abstract.cfm?URI=oe-27-21-30726},
	doi = {10.1364/OE.27.030726},
	language = {en},
	number = {21},
	urldate = {2026-01-29},
	journal = {Optics Express},
	author = {El Dirani, Houssein and Youssef, Laurene and Petit-Etienne, Camille and Kerdiles, Sebastien and Grosse, Philippe and Monat, Christelle and Pargon, Erwine and Sciancalepore, Corrado},
	month = oct,
	year = {2019},
	pages = {30726}
}

@article{liu_high-yield_2021,
	title = {High-yield, wafer-scale fabrication of ultralow-loss, dispersion-engineered silicon nitride photonic circuits},
	volume = {12},
	issn = {2041-1723},
	url = {https://www.nature.com/articles/s41467-021-21973-z},
	doi = {10.1038/s41467-021-21973-z},
	language = {en},
	number = {1},
	urldate = {2026-01-29},
	journal = {Nature Communications},
	author = {Liu, Junqiu and Huang, Guanhao and Wang, Rui Ning and He, Jijun and Raja, Arslan S. and Liu, Tianyi and Engelsen, Nils J. and Kippenberg, Tobias J.},
	month = apr,
	year = {2021},
	pages = {2236},
}

@article{pfeiffer_ultra-smooth_2018,
	title = {Ultra-smooth silicon nitride waveguides based on the {Damascene} reflow process: fabrication and loss origins},
	volume = {5},
	issn = {2334-2536},
	shorttitle = {Ultra-smooth silicon nitride waveguides based on the {Damascene} reflow process},
	url = {https://opg.optica.org/abstract.cfm?URI=optica-5-7-884},
	doi = {10.1364/OPTICA.5.000884},
	language = {en},
	number = {7},
	urldate = {2026-01-29},
	journal = {Optica},
	author = {Pfeiffer, Martin H. P. and Liu, Junqiu and Raja, Arslan S. and Morais, Tiago and Ghadiani, Bahareh and Kippenberg, Tobias J.},
	month = jul,
	year = {2018},
	pages = {884},
}

@misc{Borofloat33,
 title = {Technical {Details} of {BOROFLOAT}®},
 url = {https://www.schott.com/en-gb/products/borofloat-p1000314/technical-details},
 journal = {https://www.schott.com},
}

@article{Shah:05,
author = {Lawrence Shah and Alan Y. Arai and Shane M. Eaton and Peter R. Herman},
journal = {Opt. Express},
number = {6},
pages = {1999--2006},
publisher = {Optica Publishing Group},
title = {Waveguide writing in fused silica with a femtosecond fiber laser at 522 nm and 1 MHz repetition rate},
volume = {13},
month = {Mar},
year = {2005},
url = {https://opg.optica.org/oe/abstract.cfm?URI=oe-13-6-1999},
doi = {10.1364/OPEX.13.001999}
}

@article{chen_femtosecond-laser-written_2018,
	title = {Femtosecond-laser-written {Microstructured} {Waveguides} in {BK7} {Glass}},
	volume = {8},
	issn = {2045-2322},
	url = {https://www.nature.com/articles/s41598-018-28631-3},
	doi = {10.1038/s41598-018-28631-3},
		language = {en},
	number = {1},
	urldate = {2026-03-17},
	journal = {Scientific Reports},
	author = {Chen, George Y. and Piantedosi, Fiorina and Otten, Dale and Kang, Yvonne Qiongyue and Zhang, Wen Qi and Zhou, Xiaohong and Monro, Tanya M. and Lancaster, David G.},
	month = jul,
	year = {2018},
	pages = {10377},
}

@article{Dong:13,
author = {Ming-Ming Dong and Cheng-Wei Wang and Zheng-Xiang Wu and Yang Zhang and Huai-Hai Pan and Quan-Zhong Zhao},
journal = {Opt. Express},
number = {13},
pages = {15522--15529},
publisher = {Optica Publishing Group},
title = {Waveguides fabricated by femtosecond laser exploiting both depressed cladding and stress-induced guiding core},
volume = {21},
month = {Jul},
year = {2013},
url = {https://opg.optica.org/oe/abstract.cfm?URI=oe-21-13-15522},
doi = {10.1364/OE.21.015522}
}

@article{10.1117/1.1905363,
author = {Barry Luther-Davies and Andrei V. Rode and Nathan R. Madsen and Eugene G. Gamaly},
title = {{Picosecond high-repetition-rate pulsed laser ablation of dielectrics: the effect of energy accumulation between pulses}},
volume = {44},
journal = {Optical Engineering},
number = {5},
publisher = {SPIE},
pages = {051102},
year = {2005},
doi = {10.1117/1.1905363},
URL = {https://doi.org/10.1117/1.1905363}
}

@article{Xiao2007Losses,
  author  = {Xiao, Shijun and Khan, Maroof H. and Shen, Hao and Qi, Minghao},
  title   = {Modeling and measurement of losses in silicon-on-insulator resonators and bends},
  journal = {Optics Express},
  year    = {2007},
  volume  = {15},
  number  = {17},
  pages   = {10553--10561},
  doi     = {10.1364/OE.15.010553}
}

@article{Little1997Microring,
  author  = {Little, B. E. and Chu, S. T. and Haus, H. A. and Foresi, J. and Laine, J.-P.},
  title   = {Microring resonator channel dropping filters},
  journal = {Journal of Lightwave Technology},
  year    = {1997},
  volume  = {15},
  number  = {6},
  pages   = {998--1005},
  doi     = {10.1109/50.588673}
}

@article{Tokushima2022OFDR,
  author  = {Tokushima, Masatoshi and Ushida, Jun},
  title   = {Demonstration of in-depth analysis of silicon photonics circuits using OFDR: waveguides with grating couplers},
  journal = {Optics Letters},
  year    = {2022},
  volume  = {47},
  number  = {1},
  pages   = {162--165},
  doi     = {10.1364/OL.444876}
}

@article{Glombitza1993CFDR,
  author  = {Glombitza, U. and Brinkmeyer, E.},
  title   = {Coherent frequency-domain reflectometry for characterization of single-mode integrated-optical waveguides},
  journal = {Journal of Lightwave Technology},
  year    = {1993},
  volume  = {11},
  number  = {8},
  pages   = {1377--1384},
  doi     = {10.1109/50.254098}
}

@article{mumlyakov_void-free_2024,
	title = {Void-free upper cladding deposition process for low-loss integrated silicon nitride photonics},
	volume = {22},
	issn = {2331-7019},
	url = {https://link.aps.org/doi/10.1103/PhysRevApplied.22.054027},
	doi = {10.1103/PhysRevApplied.22.054027},
	language = {en},
	number = {5},
	urldate = {2025-04-24},
	journal = {Physical Review Applied},
	author = {Mumlyakov, Alexandr M. and Dmitriev, Nikita Yu. and Shibalov, Maksim V. and Filippov, Ivan A. and Trofimov, Igor V. and Danilin, Andrei N. and Lobanov, Valery E. and Bilenko, Igor A. and Tarkhov, Michael A.},
	month = nov,
	year = {2024},
	pages = {054027},
}

@article{Regener1985,
  author  = {Regener, R. and Sohler, W.},
  title   = {Loss in low-finesse {Ti:LiNbO3} optical waveguide resonators},
  journal = {Applied Physics B},
  volume  = {36},
  pages   = {143--147},
  year    = {1985},
  doi     = {10.1007/BF00691779}
}

@article{Walker1985,
  author  = {Walker, R. G.},
  title   = {Simple and accurate loss measurement technique for semiconductor optical waveguides},
  journal = {Electronics Letters},
  volume  = {21},
  number  = {13},
  pages   = {581--583},
  year    = {1985},
  doi     = {10.1049/el:19850411}
}

@article{Dmitriev2022JETP,
  author  = {Dmitriev, N. Yu. and Voloshin, A. S. and Kondratiev, N. M. and Lobanov, V. E. and Min'kov, K. N. and Shitikov, A. E. and Danilin, A. N. and Lonshakov, E. A. and Bilenko, I. A.},
  title   = {Measurement of Dispersion Characteristics of Integrated Optical Microresonators and Generation of Coherent Optical Frequency Combs},
  journal = {Journal of Experimental and Theoretical Physics},
  volume  = {135},
  pages   = {9--19},
  year    = {2022},
  doi     = {10.1134/S1063776122060085}
}

@article{Chang2025PolymerGlass,
  author  = {Chang, Yi and Cao, Jia-Hao and Lin, Jia-Huei and Liang, I-Chang and Wang, Cheng-Chi and Wang, Pei-Hsun},
  title   = {Low-loss polymeric waveguides for co-packaged optics on glass substrates},
  journal = {Journal of Physics: Photonics},
  volume  = {7},
  number  = {4},
  pages   = {045025},
  year    = {2025},
  doi     = {10.1088/2515-7647/ae079b}
}

@article{Jin2025SiNGlass,
  author  = {Jin, Taewon and Yoon, Seokhyeon and Jo, Kyungjin and Jung, Heeyun and Shin, Seokyoung and Lee, Jonggeon and Park, Seungwoo and Lee, Byungou and Choi, Wonmyoung and Jang, Minchul and Jung, Soo-Yong and Kim, Myunghwan and Lee, Jong Jin and Kim, Younghyun},
  title   = {Silicon nitride photonic platform on glass for scalable, high-density optical redistribution layers in panel-level packaging},
  journal = {Optics Express},
  volume  = {33},
  number  = {24},
  pages   = {50422--50431},
  year    = {2025},
  doi     = {10.1364/OE.576420}
}

@article{Brusberg2025IonExchangedGlass,
  author  = {Brusberg, Lars and Granados-Baez, Marissa and Schilling, Ryan D. and Holguin-Lerma, Jorge A. and Rodrigues, Janderson R. and Yeary, Lucas W. and Zakharian, Aramais R. and Johnson, Betsy J. and Dejneka, Matthew J.},
  title   = {Optical Design and Applications for Ion-Exchanged Glass Waveguide Circuits},
  journal = {IEEE Transactions on Components, Packaging and Manufacturing Technology},
  volume  = {15},
  number  = {8},
  pages   = {1614--1624},
  year    = {2025},
  doi     = {10.1109/TCPMT.2025.3526500}
}

@article{Hsu2025TiO2,
  author  = {Hsu, Po-Kai and Wang, Zi-Wei and Chi, Meng-Fu and Kuo, Chien-Cheng and Wang, Pei-Hsun},
  title   = {Development of high-quality TiO$_2$ photonics with E-gun evaporation},
  journal = {Optics Express},
  volume  = {33},
  number  = {16},
  pages   = {34510--34524},
  year    = {2025},
  doi     = {10.1364/OE.569771}
}

@article{polyanskiy_refractiveindexinfo_2024,
	title = {Refractiveindex.info database of optical constants},
	volume = {11},
	issn = {2052-4463},
	url = {https://doi.org/10.1038/s41597-023-02898-2},
	doi = {10.1038/s41597-023-02898-2},
	number = {1},
	journal = {Scientific Data},
	author = {Polyanskiy, Mikhail N.},
	month = jan,
	year = {2024},
	pages = {94},
}

\foreach \pdfpage in {1,...,2}{%
    \clearpage
    \includepdf[
        pages={\pdfpage},
        pagecommand={}
    ]{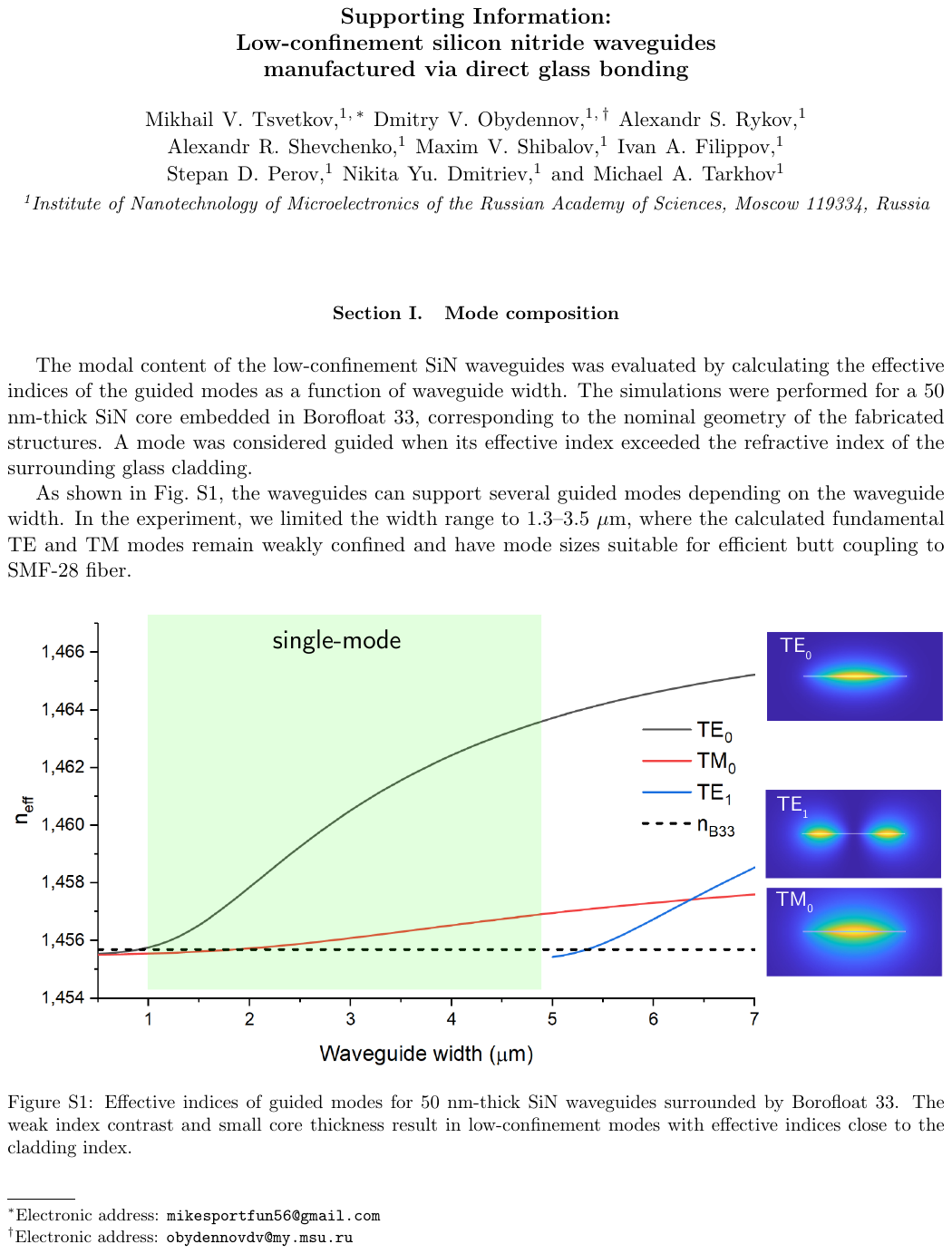}%
}

\end{document}